\documentstyle[prl,twocolumn,aps,epsfig,epsf]{revtex}
\begin{document}
\draft
\title{Metal--Insulator Transition in Randomly Interacting Systems}
\author{B. Rosenow$^1$ and R. Oppermann$^2$}
\address{$^1$Institut f\"ur Theoretische Physik. Universit\"at  zu K\"oln, 
D--50937 K\"oln}
\address{$^2$Institut f\"ur Theoretische Physik, Universit\"at W\"urzburg, 
D--97074 W\"urzburg}
\date{\today}
\maketitle
\begin{abstract}
We discuss a metal--insulator transition caused by random couplings of
magnetic moments in itinerant systems. An analytic solution for the single 
particle Green function is derived from dynamical self 
consistency equations, the corresponding density of states is characterized
by the opening of a gap. The scaling behavior of observables is analyzed in 
the framework of a scaling theory and different crossover lines are identified.
A fluctuation expansion around the mean field solution accounts for both
interaction and localization effects in a consistent manner and is argued to
be relevant for the description of the recently discovered metal--insulator 
transition \cite{Krav95,Sim97} in 2d electronic systems.
\end{abstract}
\narrowtext \tighten 
Interaction driven metal--insulator transitions (MITs) as well as magnetism 
are 
among the most striking consequences of strong electronic correlations. 
For many years progress in 
the theoretical investigation of these phenomena was 
impeded by their manifestly nonperturbative nature, 
and only the introduction of powerful new methods like self-consistent 
dynamical mean field theories (DMFT) has initiated progress 
\cite{Voll94,GeKr}.
The modification of the disorder induced localization transition by 
interactions 
has been the subject of intensive studies \cite{BeKi},  whereas the
profound effects of random interactions on the band structure have attracted 
interest only recently
\cite{Dai,RoBr} in the form of a hard gap causing a crossover from
variable range hopping to activated behavior in the conductivity.
In this letter we explore the phase diagram of metallic spin glasses in the
vicinity of the quantum phase transition between a gapped insulating phase
and a metallic phase (SG--MIT) with a density of states (DoS)
at the Fermi level increasing 
continuously from zero as the 
magnetic interaction strength is reduced. Applications to the recently 
discovered MIT in 2d systems \cite{Krav95,Sim97} and the spin glass phase in 
High $T_c$ superconductors \cite{Chou95} are discussed.\\
While in clean systems the MIT so far can only be described with the help of
numerical methods, the prevalence of disorder correlations in spin glasses
makes it possible to find an analytic solution of dynamical mean field 
equations. A Hamiltonian with hopping elements out of the Gaussian orthogonal
ensemble leads in infinite space dimensions to a set of equations 
equivalent to the local impurity self-consistent approximation (LISA) 
\cite{GeKr} as was already pointed out in \cite{RoBr}. For the typical magnetic
interaction $J$ much smaller than the kinetic energy the DoS is semicircular 
with a
width $2 E_0$ and the
system is a metallic spin liquid, whereas for $E_0= (32/3 \pi) J$ a quantum 
transition from 
paramagnet to spin glass takes place \cite{OpBi,SRO,Georges}. A further
decrease of $E_0/J$ leads to a suppression of the DoS at the Fermi level
$\rho_F$ until at 
$E_{0c}=2.59 J$
a gap opens up. The dynamical selfconsistent theory treats
both the disordered electron hopping and the spin--spin interaction in a
consistent  manner and is therefor a good starting point for a fluctuation 
expansion. From the appearance of replica symmetry 
breaking in the fluctuation theory it is conjectured that this transition is
preceded by a Griffith phase in analogy with
the ferromagnetic phase in the 
random field Ising model \cite{MeMo}. Thermal fluctuations affect the 
SG--MIT in a double way: they smear out the 
transition itself and they shift its location by reducing the magnetic 
correlations.  These effects are described by crossover lines separating 
quantum critical, insulating and metallic behavior of DoS
and transport coefficients.
While in infinite space dimensions the SG--MIT has magnetic correlations at 
its origin, in finite space dimensions localization fluctuations become 
important and the transition has to be compared with the Anderson Mott MIT
\cite{BeKi}. Both the existence of the order parameter DOS and the 
quartic critical theory
\cite{BK1} with a dangerously irrelevant  variable suggest that the 
similarity persists in fluctuation theory.
However, we expect that the detailed treatment of interaction effects in
our mean field theory and the correct choice of the ground state 
will render the renormalization group (RG) theory of 
the SG--MIT free of instabilities present in the usual sigma model approach 
\cite{Fink83,CaDiLe98} to interacting disordered systems.  \\
We study a  single band model with 
the Hamiltonian
\begin{eqnarray}
H&=& \sum_{i,j} t_{ij}c_{i,\sigma}^{\dagger} c_{j,\sigma} - 
\frac{1}{2}\sum_{i,j} J_{ij} S_i S_j\ \ ,
\label{one}
\end{eqnarray}
\\
comprised of a kinetic part with real symmetric hopping matrix elements 
distributed according to a Gaussian probability measure with moments
$<t_{ij}>=0$ and $<(t_{ij})^2>= M(i-j)$ ,  and a random interaction between 
Ising spins with variance $J^2$ (SK model). It can be viewed as effective 
$t$--$J$ model describing the interaction of spin fluctuations at
randomly distributed impurity sites with strong Coulomb repulsion 
\cite{Hertz79}.
The disorder average is performed by means of the replica method, the 
replicated 
partition function is represented with the help of Grassmann integrals.
The four fermion and four spin terms are decoupled with quaternionic matrix
fields
$\underline{\underline{R}}^{ab}(x;\tau,\tau^{\prime})$ and 
scalar fields  $Q^{ab}( x;\tau,\tau^{\prime})$.
In infinite space dimensions dynamical saddle point equations  become
exact and the stationary values can be identified with the single particle
Green function and the spin autocorrelation function, respectively. 
After further decoupling of the spin--spin interaction  the fermions can be 
integrated
out. For  a replica symmetric and static approximation of the linear 
susceptibility and in the
limit of zero temperature two of the three integrals over spin decoupling 
fields can be solved exactly by the method of steepest descent.
One ends up with a spectral type 
self--consistency equation for the the thermal single particle Green function 
$G(\epsilon_l)=(4/i E_0^2) r_l$ with $r_l$ denoting the saddle point value
of the charge decoupling fields. The shortcomings of this calculation are 
remedied in the framework of a Landau Ginzburg theory for the SG--MIT,
 which allows to include the effects of finite temperature,
inelastic interactions and replica symmetry breaking.  The
self--consistency equation for the
Green function reads
\begin{eqnarray}
r(\epsilon_l)=\frac{E_0^2}{4}(r_l + \epsilon_l)\ \ \int_{-\infty}^{\infty}
\frac{d z}{\sqrt{2\pi }} 
\frac{A(z)}{(r_l+\epsilon_l)^2+z^2}
\label{two}
\end{eqnarray}
where the weight function is given by
\begin{eqnarray}
A(z)&=&\frac{1}{\sqrt{q}}\theta[|z|\!-\! \chi_0 |\phi^{\prime}(z)|]
[1-\chi_0 \phi^{\prime \prime}(z)]
e^{-\frac{1}{2 q}[|z|- \chi_0 |\phi^{\prime}(z)|]^2}\ .
\nonumber\\
&\hspace*{-1cm}&\hspace*{-1cm}\label{three} 
\end{eqnarray}
The Edwards Anderson order parameter $q=\lim_{t\to\infty}<S(t)S(0)>$ 
and 
the zero frequency component $\chi_0$ of the local magnetic susceptibility
have to be determined by extremizing the free energy,
$\phi(z)=Tr \log[1+z^2/(r_l+\epsilon_l)^2]$ is a functional of the 
electron Green function, all energies and magnetic fields are measured
in units of the average spin coupling $J$. The derivation 
of eqs.(\ref{two},\ref{three}) will be presented elsewhere.\\
In order to gain insight into the behavior of the spectral type function 
$A(z)$ it is useful to calculate it non  selfconsistently 
for a semi-elliptic 
Green function and different values of the bandwidth $E_0$.
Large values
of $E_0$ yield a functional
$\phi(z)$ with a maximum at $z=0$ and $A(z)=A_0+O(z)$ for small z, consequently
solutions of (\ref{two}) are metallic.  For $E_0<1.5$ however, $\phi$ develops
a double peak structure and therefor  the weight function A(z) is gapped
around z=0. In this case the denominator in eq.(\ref{two}) can be expanded in
powers of $r_l+\epsilon_l$. Keeping only the leading frequency dependence
and adding the dominant temperature correction the result reads 
\begin{eqnarray}
(\delta_0 - T \delta_1) r_l + \kappa (r_l)^3= \epsilon_l
\label{four}
\end{eqnarray}
where 
\begin{eqnarray}
\delta_0&=&(\frac{4}{E_0^2<\frac{1}{z^2}>}-1),\ \ \ \ \ \ \ \ \   
\kappa=\frac{<\frac{1}{z^4}>}{<\frac{1}{z^2}>},
\label{five}
\end{eqnarray}
and $\delta_1$ is a positive constant. All expectation values are taken with 
respect to the weight function $A(z)$.
For the special value of the bare band width $E_0=2/\sqrt{
<1/z^2>}$ corresponding to $\delta_0 =0$ one finds at zero temperature a 
solution of  the form $r_l \sim sgn(\epsilon_l)
\sqrt[3]{|\epsilon_l|}$ with a dynamical critical exponent $z=3$.
In order to locate the critical point we have performed a selfconsistent 
calculation of the Green function. 
To find an analytic expression for the functional $\Phi$
the self--energy is modeled by $r_l=\kappa(\frac{ |\epsilon_l|}{\kappa})^{
1/3}$ 
for $|\epsilon_l| < \Lambda$ with a cutoff $\Lambda$ determined by
the normalization condition for the DoS $1=\int_{-\Lambda}^{\Lambda} 
\rho(\epsilon)$. A free propagator $G(\epsilon_l)=\frac{1}{i \epsilon_l}$ 
is used for
energies larger than the cutoff $\Lambda$. The functional $\Phi$ is in this
approximation given by
\begin{eqnarray}
\Phi(z)&=& \frac{1}{2 \pi} \left[ 2 \Lambda \ln\frac{1 +\frac{ 
z^2 \Lambda^{2/3}}{\kappa^{4/3}}}{1+\frac{z^2}{\Lambda^2}}
\  + 2\pi 
|z| - 4 z \arctan\frac{\Lambda}{z} \right.\\
& & \left. + \frac{4}{3} z^2\left(\frac{ 3 \Lambda^{1/3}}{
\kappa^{4/3}}
-\frac{3 z}{\kappa^2} \arctan\frac{\kappa^{2/3} \Lambda^{1/3}}{z}
\right)\right].\nonumber
\label{selfphi}
\end{eqnarray}
Using this functional we have solved numerically the system of coupled 
selfconsistency 
equations for  critical band width, local susceptibility, and spin glass order
parameter, and found the result
$E_{0c}=2.53 J$, $\chi_{0c}=0.79/J$, and
$q_c=0.66$. We have verified that a solution $\sim \sqrt{\epsilon_l}$ 
corresponding to a parabolic gap in A(z) does not satisfy 
eq(\ref{two},\ref{three}).\\
To discuss various crossover scenarios one needs the full solution for the 
single particle Green function, though. It is given by
\begin{eqnarray}
G(\epsilon_l)&=&\frac{-4 i}{E_0^2}\left\{\frac{-\delta}{3 \kappa^{2/3}[\frac{
\epsilon_l}{2}+\sqrt{\frac{\delta^3}{27 \kappa}+(\frac{\epsilon_l}{2})^2}]^{
1/3}}\right.\nonumber\\
& &\left.+\kappa^{-1/3}[\frac{
\epsilon_l}{2}+\sqrt{\frac{\delta^3}{27 \kappa}+(\frac{\epsilon_l}{2})^2}]^{
1/3}\right\} \ .
\label{six}
\end{eqnarray}
\begin{figure}
\epsfig{file=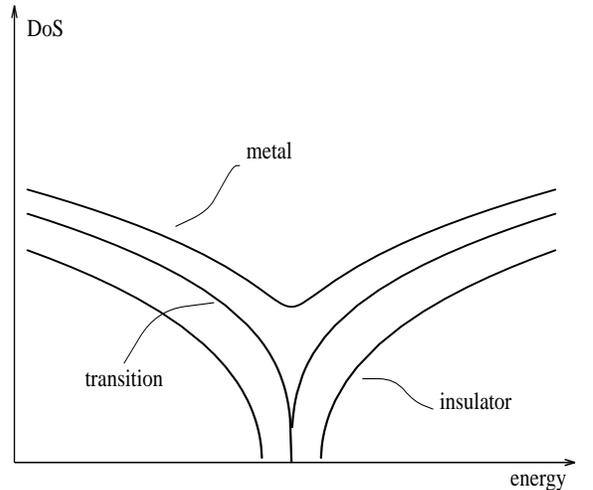,width=6.5cm,angle=270}
\caption{Single
 particle density of states DoS versus energy for different 
values of the electron hopping strength corresponding to metallic, 
critical and insulating behavior.}
\end{figure}

Here $\delta=\delta_0- T \delta_1$ plays  role of a mass squared which 
separates the disordered (insulating) phase with positive $\delta$ and 
vanishing $\rho_F$ from the ordered (metallic) phase. The dependence of 
the DoS on the bare bandwidth $E_0$ is displayed in fig.(1).
The temperature dependence of $\delta$  describes the effects
originating  in the reduction of magnetic correlations by 
thermal fluctuations.   The DoS at the Fermi level depends on temperature
via $\rho_F=\frac{4}{\pi E_0^2}\sqrt{\frac{-\delta}{\kappa}}$. 
For low T the
constant term $\delta_0$ dominates and the DoS is approximately constant,
whereas for $T \gg\frac{-\delta_0}{\delta_1}$ thermal fluctuations take over
and the system behaves as if it were at its quantum critical
point.  Hence for 
all 
observables dominated by the critical DoS, the quantum critical (QC) region is
delimited by the lines $T=\frac{|\delta_0|}{\delta_1}$, see fig.(2).\\
Given the fact that in the replica symmetric solution of the SK--model  
(as well as in simulations of finite dimensional spin glasses) 
the distribution
of local fields $P(h_{loc})$ is nonzero  for $h_{loc}=0$ it is 
questionable whether the frozen in magnetic moments 
in a spin glass can cause a gap
in the single particle DoS and drive the system insulating. 
The investigation of generalized 
Thouless--Anderson--Palmer (TAP) equations \cite{RoBr,ReOp98} suggests that  
a spin at site $i$  lowers its energy by polarizing  its neighborhood.
The resulting self energy $E_g=[\sum_j \chi_{jj} (J_{ij})^2]_{av}$ favors
the single occupation of sites and thus insulating behavior. The influence 
of replica symmetry breaking and finite range interactions on the magnetic
self energy are discussed in \cite{RoBr,RoBr98a,RoBr98b}.\\
For finite dimensional systems the DC conductivity is calculated to leading 
order in a $1/N$--expansion \cite{WeOp} from the 
density--density response function 
$D(q,\omega_n)$ utilizing the Kubo relation
\begin{eqnarray}
\sigma_{DC}=- e^2 \lim_{\omega\to 0} \omega \lim_{q\to 0}\frac{\partial}{
\partial q^2} Im D^R (q,\omega)\ \ .
\end{eqnarray}
 The result $\sigma_{DC}=e^2\rho_F D$ with
the charge diffusion constant $D=2 \pi b_m \rho_F$ ($b_M$ being the second
Fourier coefficient of the hopping correlation function M(q)) exhibits a linear
dependence on T in the QC region since there the DoS is suppressed like 
$\sqrt{T}$.\\
The results of the last paragraphs can be collected in the unifying framework
of a scaling theory. In the finite temperature formalism the 
DoS $\rho(\epsilon_l)=-\frac{\partial
\beta F}{\partial \epsilon_l}$ is the derivative of the free energy with 
respect to an external frequency $\epsilon_l$. 
 This external frequency is
analogous to an external field in the context of a magnetic phase transition,
a nonvanishing $\rho_F$ breaks the symmetry between advanced and retarded 
energies.
The transition is sharp for zero external field, whereas in the general case 
the DoS obeys the scaling relation 
$\rho \sim |\delta|^{\beta} {\cal R}_{\pm}(\frac{\epsilon}{|\delta|^{z \beta}
})$. The limiting behavior of the scaling function is $\lim_{x\to \infty}
{\cal R}_{\pm}(x)= x^{1/z}$ and $\lim_{x\to 0} {\cal R}(x) \sim\  \theta(
\mp 1)$. The two relevant scaling fields $T$ and $E_0 - E_{0c}$ are related
by a crossover exponent $\phi =1$ and hence can be  combined in a single 
variable $\delta$.
From the denominator $\omega_n + D q^2$ of the diffusive
two particle propagator one determines the scaling behavior of the diffusion
constant as 
\begin{eqnarray}
D \sim \delta^{\nu (z-2)}\ \ ,
\end{eqnarray} 
and inferring 
the violation of hyper-scaling derived from the RG analysis we finally obtain
\begin{eqnarray}
\sigma_{DC}\sim \delta^{\nu(d-\theta_t-2)}\ \ .
\end{eqnarray}
So far all results have been obtained in the framework of a replica
symmetric and spin static dynamical mean field calculation. In the
following
we will argue that the physics of a finite dimensional system with
replica
symmetry breaking and inelastic many--particle interactions is
correctly
described by a dynamical Landau--Ginzburg theory.\\
\vspace*{-.6cm}
\begin{figure}
\epsfig{file=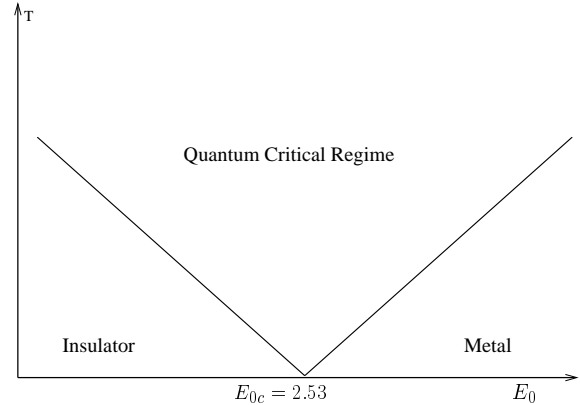,width=5.5cm,angle=270}
\caption{Phase diagram in the vicinity of the SG--MIT at
a bare bandwidth $E_0=2.53 J$. The transition lines separate
the quantum critical
region with a $T^{1/2}$--dependence of the DoS from 
the insulating regime with a gap at lower $E_0$
and from the metallic region.}
\end{figure}
In order to calculate critical properties of the {\it finite dimensional}
SG--MIT we have expanded the free energy in fluctuations about the dynamic mean field
solution. The fluctuation fields are hermitian $4n\times 4n$ matrices which
depend on two time and one space variable.  They obey charge conjugation
symmetry in the form $\underline{\underline{C}}\underline{\underline{R}}^T
\underline{\underline{C}}^T=\underline{\underline{R}}^{\dagger}$ with the
charge conjugation matrix $\underline{\underline{C}}=i \sigma_2\otimes
\sigma_0$.  We obtained for the quantum action governing the 
critical theory after a suitable rescaling of space and time variables 
\begin{eqnarray}
{\cal A}&=&\frac{1}{t}\int d^d x\{ T^2\sum_{\epsilon_l,\epsilon_k}R^{ab}_{
\alpha\beta}(x;
\epsilon_k,\epsilon_l)[(-\nabla^2+ \epsilon_k^{2/3}+\epsilon_l^{2/3}\nonumber\\
&\hspace*{-.7cm} &\hspace*{-.7cm}
+\epsilon_k^{1/3}\epsilon_l^{1/3}
+1)\delta_{\alpha\delta}\delta_{\beta\gamma}
- \gamma^z_{\alpha\delta}
\gamma^z_{\beta,\gamma}]R^{ba}_{\gamma\delta}(x;\epsilon_l,\epsilon_k)\}
\nonumber\\
& &\hspace*{-.7cm}+\frac{T^2}{t}\sum_{\epsilon_l,\epsilon_k}
R^{aa}_{\alpha\beta}(x;\epsilon_k,\epsilon_k)\gamma^z_{\beta\alpha}
r^{ab}\gamma^z_{\gamma\delta}R^{bb}_{\delta\gamma}(x;\epsilon_l,\epsilon_l)
\nonumber\\
&\hspace*{-.7cm} &\hspace*{-.7cm}
+u T^3\sum_{\epsilon_k,\epsilon_l,\epsilon_m}
R^{aa}_{\alpha\beta}(
x;\epsilon_k,\epsilon_l)\gamma^z_{
\beta\alpha}\gamma^z_{\gamma\delta}R^{aa}_{\delta\gamma}(x;\epsilon_m,
\epsilon_k-\epsilon_l+\epsilon_m)\nonumber\\
& &\hspace*{-.7cm} +\frac{\kappa_1}{4!}Tr[(\underline{\underline{R}}\underline{\underline{
\gamma^z}})^4] + \frac{\kappa_2}{4!} Tr[(\underline{\underline{R}}\underline{
\underline{\gamma^z}})^2]^2\}.
\label{MITac}
\end{eqnarray}
Here $\gamma^z=\sigma_0\otimes\sigma^z$ is a spin matrix and 
$r^{ab}$ depends on 
the Parisi solution of the spin glass problem.
The $ 4 n^2$  critical modes satisfy $\underline{\underline{R}}\underline{
\underline{\gamma^z}}=\underline{\underline{\gamma_z}}\underline{
\underline{R}}$. Note that the quantum 
interaction $u$
has no influence on the mean--field solution as for a spin glass there is no
net magnetization. The spin glass order parameter $q^{ab}$ modifies the 
random field term and describes the influence of frozen magnetic moments. 
The dangerously 
irrelevant variable t is already introduced at tree  level to account for
the shift of the upper critical dimension caused by disorder correlations.
Due to the presence of replica symmetry breaking the value of its exponent
$\theta_t$ will be different from 2 already in a two loop RG
calculation. From the presence of replica symmetry breaking the appearance
of a glassy phase preceding the SG--MIT can be conjectured in analogy to
\cite{MeMo}.\\
In a tree level RG--analysis the quantum mechanical many--body interaction $u$ is found to be irrelevant with scaling exponent $- \theta_u = -1$
in the vicinity of the upper critical dimension $d_c^{(u)}=6$. Actually $u$
is dangerously irrelevant and changes the crossover line between metallic 
and quantum critical region from the naively expected 
$T \sim \delta^{z \nu}$ to $T \sim \delta^{(z \nu)/(1 + \theta_u \nu)}$.
As long as the quantum dynamics are irrelevant 
the lower critical dimension $d_c^{(l)}$ of the model is 
 expected to be equal to
two.  However, if the interaction term becomes relevant in low spatial 
dimensions as it happens in the Finkel'stein theory of interacting electrons,
it will shift $d_c^{(l)}$ below two. This effect of a reduced 
lower critical dimension is well known in models with relevant quantum 
dynamics.
Recently a MIT has been observed in high--mobility metal--oxide--semiconductor
field--effect transistors (MOSFETs) \cite{Krav95,Sim97} and a variety of 
other 2d systems. These 
experiments challenged the widely held belief that two is the lower critical
dimension for MIT's in all types of disordered systems. The RG theory of the
MIT in interacting systems \cite{Fink83,CaDiLe98} indeed admits the 
possibility of a metallic phase but is hampered by the occurrence of a magnetic
instability in the triplet channel. A transition to a ferromagnetic state 
seems unlikely as the 2d metallic phase is quenched by a parallel 
magnetic field \cite{Pud97}. Since randomness is relevant in these materials
it is reasonable to expect spin glass freezing of magnetic moments. 
The divergence of the triplet amplitude in the RG theory  
\cite{Fink83,CaDiLe98} occurs at a finite length scale, therefor the frozen 
moments 
are most likely  not single spins but ferromagnetically ordered clusters.
These ordered clusters may form in regions with strong disorder, interact
via a RKKY type interaction mediated by conduction electrons, and form a
Stoner glass \cite{Hertz79}. The freezing of magnetic moments as represented
by the random field term in eq.(\ref{MITac}) reduces the
magnetic susceptibility and is expected to control the  runaway flow in 
a RG analysis. The validity of these ideas can be tested by performing 
magnetic measurements and looking for irreversibilities indicative of 
random freezing of magnetic moments.\\
In perovskite materials with narrow bands and a pronounced spin glass phase
like LaSrCuO on the other hand
 there should 
be a possibility to observe the interaction driven SG--MIT by lowering the
temperature below the spin glass freezing temperature. \\
In summary, we derived dynamical selfconsistency equations describing 
itinerant systems with random magnetic couplings. The analytic solution 
displays a metal--insulator transition due to magnetic correlations. The 
mean field theory served as a starting point for the derivation of a 
fluctuation theory describing the combination of localization and
interaction fluctuations. The effects of randomly frozen magnetic moments are
described by a random field term, and from the appearance  of replica symmetry 
breaking 
in this term the presence of a glassy phase preceding the MIT was 
conjectured.\\
{\it Acknowledgment}. We thank C. DiCastro and V. Pudalov for valuable 
discussions.

\end{document}